\documentclass[10pt,journal,compsoc]{IEEEtran}

\ifCLASSOPTIONcompsoc
  \usepackage[nocompress]{cite}
\else
  \usepackage{cite}
\fi
\ifCLASSINFOpdf
  \usepackage[pdftex]{graphicx}
\else
\fi
\usepackage{tikz}
\usetikzlibrary{shapes.geometric}
\newcommand\encircle[1]{%
  \tikz[baseline=(X.base)] 
    \node (X) [draw, shape=circle, inner sep=0] {\strut #1};}
\newcommand\ensquare[1]{%
  \tikz[baseline=(X.base)] 
    \node (X) [draw, regular polygon, inner sep=-1.6, regular polygon sides=4] {\strut #1};}

\begin{document}
%
\title{Amending the Characterization of\\Guidance in Visual Analytics}
%
%
%
%

\author{Davide Ceneda, Theresia Gschwandtner, Thorsten May,\\Silvia Miksch, Hans-J\"org Schulz, Marc Streit, Christian Tominski
\IEEEcompsocitemizethanks{
\IEEEcompsocthanksitem Davide Ceneda, Theresia Gschwandtner, and Silvia Miksch are with the Vienna University of Technology, Austria.\protect\\
E-mail: \{davide.ceneda,theresia.gschwandtner,silvia.miksch\}@tuwien.ac.at
\IEEEcompsocthanksitem Thorsten May is with the Fraunhofer IGD, Darmstadt, Germany.\protect\\
E-mail: thorsten.may@igd.fraunhofer.de
\IEEEcompsocthanksitem Marc Streit is with the Johannes Kepler University, Linz, Austria.\protect\\
E-mail: marc.streit@jku.at
\IEEEcompsocthanksitem Hans-J\"org Schulz and Christian Tominski are with the University of Rostock, Germany.\protect\\
E-mail: \{hans-joerg.schulz,christian.tominski\}@uni-rostock.de}
}

\IEEEtitleabstractindextext{%
\begin{abstract}
At VAST 2016, a characterization of guidance has been presented. It includes a definition of guidance and a model of guidance based on van Wijk's model of visualization. This note amends the original characterization of guidance in two aspects. First, we provide a clarification of what guidance actually is (and is not). Second, we insert into the model a conceptually relevant link that was missing in the original version.
\end{abstract}

\begin{IEEEkeywords}
Visual analytics, guidance, assistance, user support.
\end{IEEEkeywords}}

\maketitle

\IEEEdisplaynontitleabstractindextext

%
\IEEEpeerreviewmaketitle

\ifCLASSOPTIONcompsoc
\IEEEraisesectionheading{\section{Introduction}\label{sec:introduction}}
\else
\section{Introduction}
\label{sec:introduction}
\fi

%
%
%
%
\IEEEPARstart{V}{isual} analytics (VA) is a sense-making and problem-solving technology that involves diverse analytic, visual, and interactive tools. As problems get increasingly complex, the corresponding techniques and workflows become more and more complex as well. This can lead to situations where users feel lost or do not know how to continue. The analytic progress is stalled.

At this point, it would be good to have a mechanism that can provide some support to overcome the stall. This is what guidance aims to achieve: The goal is to help the user in making continuous progress.

\section{What is Guidance and What is it not?}

In the context of VA, guidance has been described as a strategy to assist in data exploration and analysis. The original definition is as follows~\cite{Ceneda17Guidance}:

\begin{quote}
\textbf{Guidance} is a computer-assisted \emph{process} that aims to actively resolve a \emph{knowledge gap} encountered by users during an \emph{interactive} visual analytics session.
\end{quote}

The three important aspects of this definition are emphasized in italics. First, guidance is a dynamic process that runs alongside the regular data analysis activities of the user. Second, there is a knowledge gap that causes the data analysis to stall. The user does not know how to proceed. The goal of guidance is to narrow the knowledge gap. Finally, the definition of guidance describes an interactive scenario. That is, guidance assumes the existence of a human in the loop.

This original definition of guidance in VA captures the most relevant characteristics. Yet, in discussions with colleagues, we realized that a clarification would help to better understand what guidance is, and what it is not.

To clarify, guidance provides one or multiple \emph{suggestions} to the user. Suggestions can be considered or ignored by the user. Suggestions are to help users in forming decisions. Making the decisions remains the responsibility of the user.

Guidance does not aim to close the knowledge gap automatically with a definite or exact answer. Typically, this is not even possible due to ill-defined or too complex problems. If guidance were able to compute a precise answer, we could neglect VA at all, compute the answer, and provide it to the user right away. But this would contradict with the idea of the human in the loop.

In this sense, guidance is comparable to a mentor helping a student. While the mentor does not know the solution of the student's problem, he or she can provide hints as how to approach the problem, guiding the student towards finding the solution on his or her own.

It is clear now that guidance is not merely an additional algorithm that computes results, but is indeed a catalyst for human-computer cooperation.


\begin{figure*}[t]
\centering
\includegraphics[width = \linewidth]{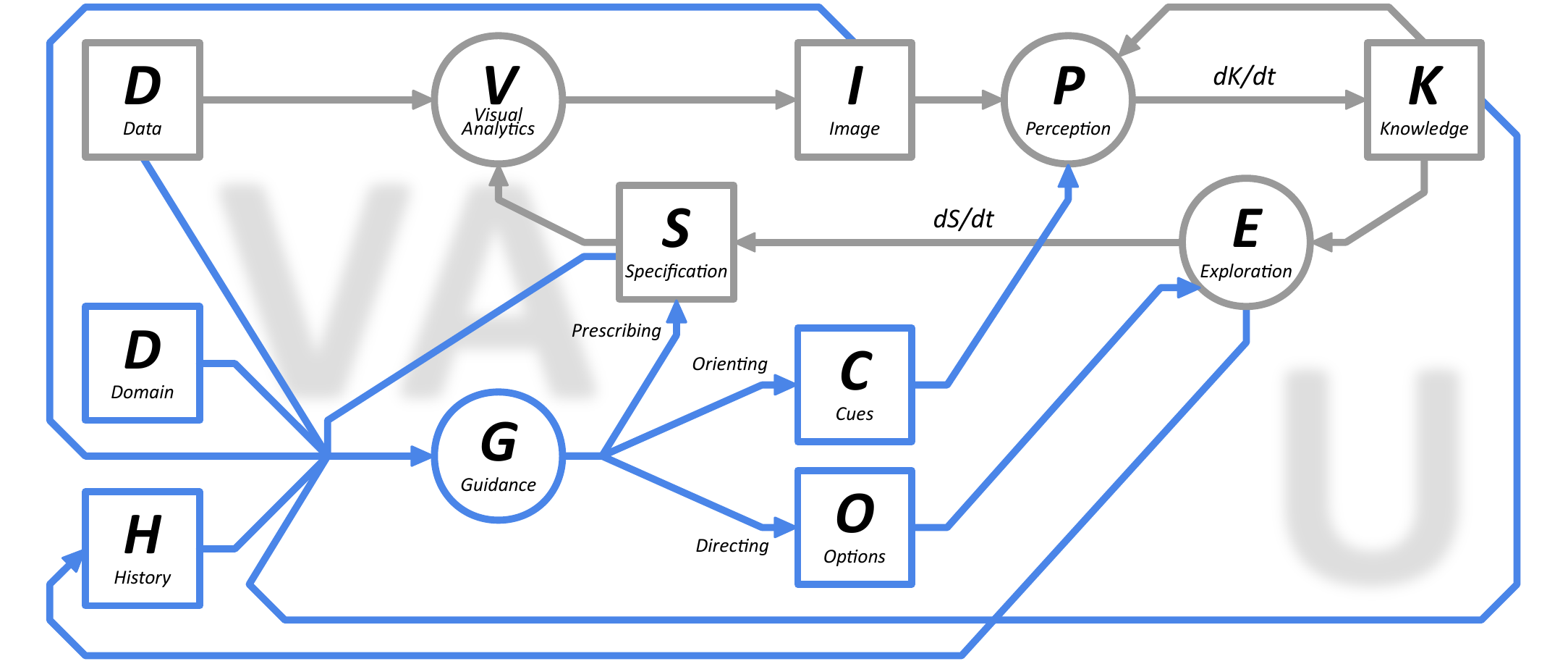}
\caption{Components of guidance (in blue) attached to van Wijk's~\cite{vanWijk06VisModel} model (in gray). Aspects of visual analytics (VA) are shown to the left, while user aspects (U) are on the right. Guidance considers the user's knowledge (or lack thereof) and may build upon various inputs, including data, interaction history, domain conventions, VA specifications, and visualization images. Different degrees of guidance are possible. Orienting uses visual cues to enhance perception. Directing supports exploration by providing alternative options. Prescribing directly operates on the specification.}
\label{fig:guidance-model}
\end{figure*}

\section{The Missing Link}

In order to explain how guidance and VA can be connected, van Wijk's model of visualization~\cite{vanWijk06VisModel} has been extended by guidance-related components~\cite{Ceneda17Guidance}. Fig.~\ref{fig:guidance-model} shows artifacts as boxes and functions as circles. They are connected by well-defined links to indicate that specific functions process some input artifacts and generate some output artifacts. Van Wijk's original model (in gray) describes how data become knowledge through perception and interactive exploration. The guidance-related extensions (in blue) are to illustrate that various inputs may be utilized by the guidance function to provide assistance in different ways.

During the presentation of our model at the VAST conference, the audience discovered that a link was missing from the VA specification \ensquare{S} to the guidance function \encircle{G}. It turned out that the missing link suggested to the audience that the VA specification (including visualization techniques, analytic algorithms, interactive tools, and their corresponding parameterization) may \emph{not} serve as an input for the guidance function. Of course, this is not correct, and indeed the missing link is a glaring omission in the guidance model.

Therefore, Fig.~\ref{fig:guidance-model} provides an updated version of the model that includes the missing link from \ensquare{S} to \encircle{G}. This makes much more sense, and it avoids any potential misinterpretation due to the absence of that link in the original model.

In fact, the model is now more appropriate in that any artifact can be used as an input to provide guidance. For instance, the data \ensquare{D} that are analyzed or the history \ensquare{H} of the interactions carried out so far can be input to the guidance function. Also the images \ensquare{I} that are generated through VA can serve as input. For example, Wattenberg and Fisher study the perceptual organization of visual representations~\cite{Wattenberg03Perceptual}, which could potentially be useful to derive hints for guidance. If it is possible to use images \ensquare{I} as input, why should it not be possible to utilize the specification \ensquare{S} of the image generation, which was what the missing link suggested.

This gap is closed now. Guidance can of course use the VA specification as an input. An example would be to look at the current parameterization of analytic computations and suggest that alternative settings could lead to new insights that would narrow the knowledge gap. One could also look at the specification of the visual representation. For example, there is already existing work that deconstructs and restyles visualizations that are specified via D3 scripts~\cite{Harper14RestylingD3}. A similar method could serve as a basis to analyze the specification and adjust it in order to integrate visual cues that provide guidance.

\section{Conclusion}

This note briefly commented on two amendments of the definition and model of guidance in VA as provided in previous work~\cite{Ceneda17Guidance}. The definition has been clarified to clearly specify guidance as a mechanism that may offer suggestions, but no definite answers. The model has been amended by adding a link from the VA specification to the guidance function, indicating that the VA specification is a valid and relevant input for guidance.

With these two amendments, our understanding of guidance has moved but a step forward. More substantial progress can be expected when addressing the topics for future work discussed in~\cite{Ceneda17Guidance}.


%



\ifCLASSOPTIONcompsoc
  \section*{Acknowledgments}
\else
  \section*{Acknowledgment}
\fi

The authors would like to thank the participants of the VAST session where the guidance definition and model were presented. In particular, we thank Ben Shneiderman, who asked for clarification of the role of guidance and the role of the user, and Helwig Hauser, who raised the concern about the missing link between the VA specification and the guidance function.

\ifCLASSOPTIONcaptionsoff
  \newpage
\fi



\bibliographystyle{IEEEtran}
\bibliography{GuidanceComment}
\end{document}